\documentclass[aps,prd,preprint,showpacs,superscriptaddress]{revtex4-1}
\usepackage{amsmath}
\usepackage{latexsym}
\usepackage{amsfonts}
\usepackage{graphicx}
\usepackage{mathrsfs}
\usepackage{epstopdf}
\usepackage{color}
\usepackage{bm}
\usepackage[colorlinks,linkcolor=magenta,anchorcolor=blue,citecolor=blue]{hyperref}

\def\be{\begin{equation}}
\def\ee{\end{equation}}
\def\bea{\begin{eqnarray}}
\def\eea{\end{eqnarray}}

\begin{document}

\title{Reconstruction of inflation model from tensor-to-scalar ratio}

\author{Jun Chen}
\email{junchen@mails.ccnu.edu.cn}
\affiliation{Key Laboratory of Quark and Lepton Physics (MOE), Central China Normal University, Wuhan, Hubei 430079, China}
\affiliation{Institute of Particle Physics, Central China Normal University, Wuhan, Hubei 430079, China}
\affiliation{Institute of Astrophysics, Central China Normal University, Wuhan, Hubei 430079, China}

\begin{abstract}
we reconstruct the potential of minimally coupling inflation model which produces small value of the tensor-scalar ratio. In these inflation models, tensor-scalar ratio  is proportional to $\frac{1}{N^2}$ with scalar spectral index is in good agreement with recent cosmological observations, in which $N$ is the e-folding number and $p$ is a positive number. We also reconstruct $F(R)$ gravity inflation from small tensor-tensor ratio.
\end{abstract}

\maketitle

%%%  Introduction
%%%%%%%%%%%%%%%%%%%%%%%%%%%%%%%%%%%%%%%%%%%%%%%%%%%%%%%%%%%

\section{Introduction}

The cosmology is one of the pillars in the modern physics and gives the successful description of the Universe. 
The inflation theory, an essential ingredient of the modern cosmology, shows the good agreement with the current observational data of the early Universe,
and various measurements have been discussed to distinguish the models of inflation.
Observational technologies have been developing year by year, 
and we expect more precise and improved data by the forthcoming observations with high level of detectability. 

The primordial gravitational wave is a useful measurement to constraint the inflation models.
We can characterize the amplitude of the primordial gravitational wave in terms of the tensor-to-scalar ratio $r$,
which is the ratio of the amplitudes of tensor and scalar fluctuations, 
and it bring us the tight constraint on the inflation potential.
For example, the PLANCK 2015 data has put stringent constraints on $r$. From the newly-released PLANCK 2018 \cite{Akrami:2018odb}, the constraints on these parameters are $\ln(10^{10}A_s)=3.044\pm0.0014$ ($68\%$ C.L.), $n_s=0.9649\pm0.0042$ ($68\%$ C.L.), (TT, TE, EE+lowE+lensing), $r_{0.002}<0.064$ ($95\%$ C.L., TT, TE, EE+lowE+lensing+BK14).

Over the next few years, 
there will be lots of world-class projects for the detection of the low-frequency GW, such as eLISA Satellite \cite{AmaroSeoane:2012km}, TianQin Satellite \cite{Luo:2015ght}, and AliCPT Telescope \cite{Li:2017drr}. 
The AliCPT telescope (first stage), located in the Ali region of Tibet, China with an altitude of 5,250 meters, is ground-based observatory for the B-mode polarization of CMB and primordial gravitational waves, 
and it is expected to improve the current constraint on the tensor-to-scalar ratio to about one order of magnitude within a few years.
For search of the small tensor-to-scalar ratio by AliCPT, 
it is worth studying the new inflation theory which has unique mechanism to suppress the tensor-to-scalar ratio rather than the existing theories.

As an illustration, we look at the Starobinsky model which predicts the small tensor-to-scalar ratio.
In the Starobinsky model, 
we find (1) the tensor-to-scalar ratio $r$ is proportional the square of the slow-roll parameter $\epsilon$, $r\propto \epsilon^2$,
and (2) the slow-roll parameter is proportional to the inverse of e-folding number $N$, $\epsilon \propto 1/N$.
Finally, we obtain $r \propto 1/N^2$.
On the other hand, the minimally coupling inflation model with a canonical scalar field gives  $r \propto 1/N$
because the tensor-to-scalar ratio is proportional to the slow-roll parameter $r\propto \epsilon$, 
and the slow-roll parameter is proportional to the inverse of e-folding number $\epsilon \propto 1/N$.

It is notable that the Starobinsky model can be written in terms of the minimally-coupling scalar field inflation by the Weyl transformation.
The Starobinsky model is one of the $f(R)$ gravity,
and the $f(R)$ gravity is equivalent to the general relativity minimally coupling with scalar field
when one transforms the metric $g_{\mu \nu}$ as $\tilde{g}_{\mu \nu} = \Omega(x)^{2} g_{\mu \nu}$,
where $g_{\mu \nu}$ and $\tilde{g}_{\mu \nu}$ denote the metric in the Jordan frame and Einstein frame, respectively.
Thus, one can find that the relations, $r\propto \epsilon^2$ and $\epsilon \propto 1/N$, in the Jordan frame are translated as 
$r\propto \epsilon$ and $\epsilon \propto 1/N^{2}$ in the Einstein frame.

The above observation suggests that 
we may realize the situation that the tensor-to-scalar ratio is proportional to the inverse-square of e-folding number $r\propto 1/N^{2}$
if we appropriately choose (i) function of $f(R)$ or (ii) inflaton potential in minimally-coupling scalar field inflation.
In this work, we study the $f(R)$ gravity theory for the inflation based on the reconstruction method \cite{reconstruction}.
We formulate small tensor-to-scalar ratio $r$ for the generic function and specify it so that $r \propto 1/N^{2}$.
We also address the description of the scalar field inflation corresponding to the reconstructed $f(R)$ function.

Our paper is organized as follows: in Sec. \ref{ii} we briefly review the relationship between the aciton in Jordan frame and the potential in Einstein frame. In Sec. \ref{iii} we restore the potential of minimally coupling inflation. In Sec. \ref{iiii} we restore the action of pure gravity inflationary model. Sec. \ref{iiiii} attributes to concluding remarks.

\section{$f(R)$ from tensor-to-scalar ratio: Analysis in the Einstein Frame}

We start with the action in $f(R)$ gravity 
\be\label{action}
S=\frac{1}{2\kappa^2}\int d^{4}x\sqrt{-g}f(R)~,
\ee
where $\kappa^2=8\pi G=\frac{1}{M_{pl}^2}$ and also $M_{pl}$ is the planck mass.
\label{ii}
The action Eq.(\ref{action}) can be rewritten in the form
\be
\label{actionform}
S=\int d^4 x \sqrt{-g}\Big{(} \frac{1}{2\kappa^2}FR-U \Big{)}~,
\ee
where
\be
\label{U}
U=\frac{FR-f}{2\kappa^2}~.
\ee
Then, by the conformal transformation $g_{\mu\nu}=g^E_{\mu\nu}\Omega^{-2}$, the action \ref{actionform}) is transfromed as
%%%
\be
\label{actionEinsteinframe}
S=\int d^{4}x\sqrt{-g_{E}}[\frac{1}{2\kappa^{2}}R_{E}-\frac{1}{2}(\nabla_{E}\phi)^{2}-U(\phi)]~,
\ee
where the quatity with subscript $E$ denotes the one which is in Einstein frame. $\phi$ and $U(\phi)$ are given by 
\begin{align}
\label{PhiE}
\kappa\phi =&\sqrt{\frac{3}{2}}\ln f_{R}\equiv\sqrt{\frac{3}{2}}\ln F \, , \\
\label{UE}
U(\phi) =&\frac{FR-f}{2F^2}~.
\end{align}
If the potential is given, $R$ and $f(R)$ is are determined by 
\bea
\label{RV}
R&=&e^{\sqrt{\frac{2}{3}}\kappa\phi}(4\kappa^{2}U+\sqrt{6}\kappa U_{\phi}) ~,\\
\label{fRV}
f(R)&=&F^{2}(2\kappa^{2}U+\sqrt{6}\kappa U_{\phi})~.
\eea

\section{Reconstructing minimally coupling inflation potential from tensor-to-scalar ratio}
\label{iii}
We explain the method to reconstruct $V(\phi)$ for a given tensor-to-scalar ratio in the framework of single field with canonical kinetic term. The relation between e-folding number and potential is
\be
\label{efoldingpotential}
N(\phi_{\ast})=\int_{t_{i}}^{t_{f}}Hdt=\int_{\phi_{\ast}}^{\phi_{e}}\frac{H}{\dot{\phi}}d\phi\approx-\int_{\phi_{\ast}}^{\phi_{e}}\frac{1}{M_{pl}^{2}}\frac{V}{V_{\phi}}d\phi~,
\ee
in which $t_\ast$ and $\phi_\ast$ is the time and the value of field when inflation starts,  $t_e$ and $\phi_e$ is the time and the value of field when inflation ends. $3H\dot{\phi}\approx-V^\prime$ is used in the third equality.
The potential-based slow-roll parameter is 
\be
\label{SRPpotential}
\epsilon\simeq\frac{1}{2\kappa^{2}}(\frac{V_{\phi}}{V})^{2}~.
\ee

The method is based on the assumption that the tensor-to-scalar ratio has the following form
\be
\label{epsilonefolding}
\epsilon=\frac{p}{N^{2}}~,
\ee
in which $p$ is a positive number. We define Following functions
\bea
\label{fphi}
	f(\phi)&\equiv&\frac{V}{V_{\phi}}~,\\
\label{Fphi}
    F(\phi)&\equiv&\int f(\phi)d\phi~.
\eea
From Eq.(\ref{efoldingpotential}), Eq.(\ref{fphi}) and Eq.(\ref{Fphi}), we obtain that
\be
\label{NF}
N(\phi_{\ast})=\int_{\phi_{e}}^{\phi_{\ast}}\frac{V}{V_{\phi}}d\phi=\int_{\phi_{e}}^{\phi_{\ast}}f(\phi)d\phi=F(\phi_{\ast})-F(\phi_{e})~.
\ee 
Usually we can kown the value of inflation field by using the relation that $\epsilon(\phi_{e})=1$, so we can get the value of $F(\phi_{e})$, so we can define that
\be
\label{CF}
C\equiv F(\phi_{e})
\ee
when the inflation starts, the value of slow-roll parameter is
\be
\label{epsilonN}
\epsilon_{\ast}\simeq \frac{1}{2}(\frac{V_{\phi}}{V})^{2}\big|_{\phi=\phi_{\ast}}=\frac{1}{2}\frac{1}{f^{2}(\phi_{\ast})}=\frac{p}{N^{2}}~.
\ee
By combining Eq.(\ref{NF}) and Eq.(\ref{epsilonN}), we can get the following differential equation 
\be
\label{diffequation}
f=\sqrt{\frac{1}{2p}}[F(\phi_{\ast})-C] 
\ee
The solution to this differential equation is another differential equation related to potential
\be
\label{diffV}
\frac{V}{V_{\phi}}=f(\phi)=\sqrt{\frac{1}{2p}}C_{2}e^{\sqrt{\frac{1}{2p}}\phi}~,
\ee
in which $C_1$ is a positive integration number. So we can get the form of potential 
\be
\label{solutionV}
V=C_{2}e^{-\frac{1}{C_{1}}e^{-\sqrt{\frac{1}{2p}}\phi}}
\ee
in which $C_2$ is a positive integration number. We can get the spectral index of scalar perturbation 
\be
\label{sindex}
n_s-1=-2\epsilon-\epsilon \propto-a\frac{1}{N}-b\frac{1}{N^2}~,
\ee
where $a$ is a postive constant. So the spectral index of scalar perturbation based on the potential (\ref{solutionV}) is in good agreement with observations.

Supposing that potential (\ref{solutionV}) is big field potential. then $\phi$ is large, we can get
\be
\label{approxV}
V\approx C_{2}(1-\frac{1}{C_{1}}e^{-\sqrt{\frac{1}{2p}}\phi})~.
\ee
When $p=\frac{3}{4}$ and $C_2=\frac{3M^2}{4}$, we can recover the potential form of  Starobinsky’s model as a specific case $(n = 2)$ in Einstein frame \cite{fRtheories}.

For the potential (\ref{solutionV}), from Eq. (\ref{RV}) and Eq. (\ref{fRV}), we obtain 
\be
\label{fRphi}
f(R)=2C_{2}F^{2}[1+\frac{1}{C_{1}}(\sqrt{\frac{3}{4p}}-1)F^{-\sqrt{\frac{3}{4p}}}]~.
\ee

\section{Inflation in $f (R)$ Theories}
\label{iiii}
The field equation of $f (R)$ gravity can be derived by varying the action (\ref{action}) with respect to the metric $g_{\mu\nu}$ 
\be\label{fieldequation}
F(R)R_{\mu\nu}-\frac{1}{2}f(R)g_{\mu\nu}-\nabla_\mu\nabla_\nu F(R)+g_{\mu\nu}\square F(R)=0~,
\ee
where $F(R)\equiv \partial f/\partial R$, we consider the FRW space time with a time-dependent scale factor $a(t)$ and a metric 
\be\label{FRWmetric}
ds^2=g_{\mu\nu}dx^\mu dx^\nu=-dt^2+a^2(t)d{\bm{x}}^2~.
\ee
For this metric, the background equation is 
\bea
\label{rho}
3FH^2&=&(FR-f)/2-3H\dot{F}~,\\
\label{p}
-2F\dot{H}&=&\ddot{F}-H\dot{F}~.
\eea
When the action (\ref{action}) is used to describe a quite general class of inflationary theories, the slow-roll indices for the action (\ref{action}) have the following general form
\be
\label{srp}
\epsilon_1=-\frac{\dot{H}}{H^2}~,~\epsilon_2=\frac{\dot{F}}{2HF}~,
\ee
and the corresponding tensor-to-scalar ratio is \cite{f(R)review}
\be
\label{r}
r\approx 48\epsilon_2^2=12\frac{{\dot{F}}^2}{H^2 F^2}~.
\ee

Inspired by the work \cite{ratioefolding}, we assume that the tensor-to-scalar ratio is proportional to e-foldings number $N$ to the negative second power, we set
\be
\label{rationefolding}
r=\frac{q}{N^2}~,
\ee
in which q is a positive number. The e-folding number $N$ is defined by 
\be
\label{efolding}
N=\int^{t_{end}}_{t}Hdt~.
\ee
From $dN=Hdt$, the slow-roll parameter $\epsilon_2$ can be rewritten as 
\be
\label{rewrittensrp}
\epsilon_2=12\frac{\dot{F}^2}{H^2 F^2}=\frac{12(\partial F/\partial N)^2}{F^2}=\frac{12 F_N^2}{F^2}
\ee
From Eqs. (\ref{rationefolding}) and (\ref{rewrittensrp}), we obtain the relation 
\be
\label{FN}
\frac{F_N}{F}=\frac{\beta}{N}
\ee
in which $\beta=\sqrt{q/12}$. For convenience, let`s chose that $\beta=1$. The solution to the equation above is 
\be
\label{FequalN}
F=\alpha N~,
\ee
$\alpha$ is a integration constant. 
\be
\label{dotF}
\dot{F}=\frac{dF}{dt}=\frac{dF}{dN} \frac{dN}{dt}= H\frac{dF}{dN}=\alpha H~.
\ee
Subsititude the equation above into Eq (\ref{rho}) and use the approximation that $R=6H^{2}(2-\epsilon_{1})\approx 12H^2$, we get a differential equation 
\be
\label{rhodiff}
f\approx\frac{1}{2}FR+\frac{1}{2}\alpha R~.
\ee
The equation above can be solved analytically, and the solution is 
\be
\label{thodiffsolution}
f=\alpha R+CR^{2}
\ee 
in which $C$ is a integration constant. So we can restore Starobinsky model \cite{the Starobinsky model} perfectly

When $\beta\ne 1$, Eq. (\ref{FN}) is easily integrated to give 
\be
\label{FNbeta}
F(N)=\alpha N^\beta~.
\ee
Since $\dot{F}=-\alpha \beta H N^{\beta-1}=\alpha\beta H(\frac{F}{\alpha})^{1-\frac{1}{\beta}}$ and $R=6H^2(2-\epsilon_1)\approx12H^2$, Eq. (\ref{rho}) becomes 
\be
\label{rhoN}
f=\frac{1}{2}FR-\alpha\beta H^{2}(\frac{F}{\alpha})^{1-\frac{1}{\beta}}~.
\ee
This equation is known as d`Alembert`s differential equation, and the general solution is given by 
\bea
\label{rhoNsoluton}
R&=&A\exp[\int\frac{1-\frac{1}{6}(\beta-1)(\frac{F}{\alpha})^{-1/\beta}}{F+\frac{1}{6}\alpha\beta(\frac{F}{\alpha})^{1-1/\beta}}dF]\\
&=&A\left(\beta+6\left(F/\alpha\right)^{1/\beta}\right)^{(2\beta-1)}F^{(1-\beta)/\beta}
\eea
Substitude this solution into Eq. (\ref{rhoN}), $f(E)$ is given by 
\be
\label{fFrelation}
f=\frac{1}{2\alpha^{2}}6^{(2\beta-1)}AF^{2}(1-\frac{1}{2}\beta(\frac{F}{\alpha})^{-\frac{1}{\beta}})
\ee
We find that by setting $\frac{1}{2}\beta\alpha^{\frac{1}{\beta}} \rightarrow \frac{1}{C_{1}}(\sqrt{\frac{3}{4p}}-1)$ and $\frac{1}{\beta} \rightarrow \sqrt{\frac{3}{4p}}$. Eq.(\ref{fFrelation}) agrees with Eq.(\ref{fRphi}). 

\section{Summary}
\label{iiiii}
In this paper, motivated by the relation $r\propto \frac{1}{N^2}$ indicated by Starobinsky model, we reconstructed the potential of minimally coupling inflation. After we using big field potential assumption, we can recover the potential of Starobinsky`s model in Einstein frame. For the potential (\ref{solutionV}), we derived  the Langrangian of $f(R)$ by using conformal transformation.  

We also reconstructed the lagrangian of $f(R)$ gravity. After we chose appropriate integration constant, Eq.(\ref{fFrelation}) agrees with Eq.(\ref{fRphi}), So we can say that these two methods of reconstruction are equivalent.

%%%  Appendix
%%%%%%%%%%%%%%%%%%%%%%%%%%%%%%%%%%%%%%%%%%%%%%%%%%%%%%%%%%%

\end{document}